\documentclass[9pt,twocolumn,twoside]{osajnl}
\usepackage{amsmath,amsfonts}
\newcommand{\dif}{\mathrm{d}}

\journal{ol} 

\setboolean{shortarticle}{true}

\title{Tunable self-similar Bessel-like beams of arbitrary order}

\author[1]{Michael Goutsoulas}
\author[2,3]{Domenico Bongiovanni}
\author[2]{Denghui Li}
\author[2,4]{Zhigang Chen}
\author[1,2,5,*]{Nikolaos K. Efremidis}

\affil[1]{Department of Mathematics and Applied Mathematics, University of Crete, 70013 Heraklion, Crete, Greece}
\affil[2]{The MOE Key Laboratory of Weak-Light Nonlinear Photonics, TEDA Applied Physics Institute and School of Physics, Nankai University, Tianjin 300457, China}
\affil[3]{INRS-EMT, 1650 Blvd. Lionel-Boulet, Varennes, Quebec J3X 1S2, Canada}
\affil[4]{Department of Physics and Astronomy, San Francisco State University, San Francisco, California 94132, USA}
\affil[5]{Institute of Applied and Computational Mathematics, Foundation for Research and Technology - Hellas (FORTH), 70013 Heraklion, Crete, Greece}

\affil[*]{Corresponding author: nefrem@uoc.gr}

\begin{abstract}
  We predict that Bessel-like beams of arbitrary integer order can exhibit a tunable self-similar behavior (that take an invariant form under suitable stretching transformations). Specifically, by engineering the amplitude and the phase on the input plane in real space, we show that it is possible to generate higher-order vortex Bessel-like beams with fully controllable radius of the hollow core and maximum intensity during propagation. In addition, using a similar approach, we show that it is also possible to generate zeroth order Bessel-like beams with controllable beam width and maximum intensity. Our numerical results are in excellent agreement with our theoretical predictions. 
 
\end{abstract}

\setboolean{displaycopyright}{true}
\usepackage{amsmath}
\usepackage{natbib}
\DeclareMathOperator{\sech}{sech}
\DeclareMathOperator{\sgn}{sgn}

\begin{document}

\maketitle

Diffraction is a phenomenon that naturally occurs during optical wave propagation. Propagation invariant (PI) fields, in their ideal form, carry infinite power and maintain an invariant profile as they propagate~\cite{efrem-optica2019,mcglo-cp2005}. Finite power truncations of PI fields, which carry finite power and thus are experimentally realizable, can retain many of the features of ideal PI fields. The main characteristic of PI fields that is utilized in most of the applications is the high intensity lobe which, in finite power truncations, can still remain almost invariant during propagation for several diffraction lengths. There are two main classes of PI optical fields. The first is the Bessel beam~\cite{durni-prl1987,durni-josaa1987}, which relies on the generation of an extended focal line and its main lobe propagates along the optical axis. The second is the Airy beam~\cite{sivil-ol2007,sivil-prl2007}, which relies on the presence of a caustic or a catastrophe, and propagates along a parabolic trajectory. Several applications benefit from the use of PI beams, ranging from particle manipulation~\cite{arlt-oc2001,baumg-np2008,zhang-ol2011}, microscopy and imaging~\cite{planc-nm2011,jia-np2014,vette-nm2014}, filamentation~\cite{polyn-science2009}, and free space optical communications~\cite{birch-josaa2015} to mention a few.

Higher order Bessel beams exhibit a zero on-axis intensity surrounded by concentric rings. This dark spot is an immediate consequence of the phase singularity which is expressed through the azimuthial phase term $e^{i n\theta}$, and is associated with the order $n$ and thus the induced vorticity. Higher order Bessel beams carrying orbital angular momentum were first realized experimentally in~\cite{vasar-josaa1989}, but later works achieved higher efficiency by using an axicon~\cite{pater-oc1996-1} and other techniques~\cite{vette-lpr2019}. This led to exciting applications in the areas of plasma generation and filamentation~\cite{fan-pre2000}, particle manipulation~\cite{volke-job2002,zhao-sr2015}, material processing~\cite{mathi-apl2012}, and free-space optical communications~\cite{mphut-ao2019}.

Engineering the properties of Bessel beams has been an issue of fundamental importance in terms of applications. In this respect, several works have considered how to engineer the trajectory of Bessel beams. Specifically, the idea of snaking a beam was proposed in~\cite{rosen-ol1995}. Spiraling Bessel beams were proposed and observed in~\cite{jarut-ol2009,matij-oe2010}. Similar principles were used in~\cite{morri-jo2010} for a snaking beam capable of propagating around obstacles. Helicon beams result from the superposition of standard Bessel beams~\cite{pater-oc1996-2,alonz-oe2005,vetter-prl2014}. A generic approach that addresses the problem of generating a Bessel beam that follows, not particular classes of paths, but generic arbitrary trajectories was proposed in~\cite{chrem-ol2012-bessel}. It was followed by an experimental observation~\cite{zhao-ol2013} and generalizations in the non-paraxial domain~\cite{chrem-pra2013} and in the case of vortex Bessel-like beams~\cite{zhao-sr2015}. In~\cite{cizma-oe2009} a technique was proposed to engineer the axial profile of Bessel beams (axial intensity and lateral cross section) in the Fourier space. Pin beams, a class of Bessel-like beams with engineered width that decreases with the propagation distance, exhibit robust propagation through atmospheric turbulence over kilometric distances~\cite{zhang-aplp2019}.

Here we propose a method for the generation of vortex Bessel-like beams with the tunable parameters being the hollow core radius and the maximum amplitude.
This is achieved by engineering the amplitude and the phase of the optical wave on the input plane in real space.
Our method is also applied in the case of zeroth order Bessel-like beams, in which case the width as well as the axial (maximum) intensity are fully controllable.
The optical waves considered here take an invariant form under suitable stretching transformations and, thus, they can be considered as self-similar.
Our theoretical results are in excellent agreement with direct numerical simulations.

Let us start by considering the Fresnel diffraction integral for the dynamics of an optical beam in a dielectric medium in cylindrical coordinates
\begin{equation}
\psi(r,\theta,z)=
\frac{e^{i\frac{kr^2}{2z}}e^{in\theta}}{i\lambda z} 
\int_{\mathbb R} \dif\rho\int_0^{2\pi} \dif s
A(\rho)\rho e^{i\phi-ins+ik\frac{\rho^2- 2r\rho\cos s}{2z}},
\label{eq:Fresnel}
\end{equation}
where $k=2\pi/\lambda$ is the wavenumber, $\lambda$ is the optical wavelength, $(r,\theta,z)$ are cylindrical coordinates with $(r,\theta)$ being the transverse polar parameters and $z$ being the longitudinal distance, $(\rho,\xi)$ are the polar coordinates on the input plane, and $s=\theta-\xi$. In Eq.~(\ref{eq:Fresnel}) the initial condition is decomposed into amplitude and phase as $\psi_0=A(\rho)e^{i\phi(\rho)+in\xi}$, where $n$ is the topological charge. Let us first focus in the case of beams without vorticity $n=0$.
By making the assumptions that (i) the amplitude $A(\rho)$ is a slowly varying function, (ii) a single ray emerges from each radial location $\rho$, and (iii) $r\ll z/(k\rho)$, we derive the ray equation
\begin{equation}
  \phi'(\rho) = -k\rho/z
  \label{eq:j0:phip}
\end{equation}
and the following relation for the dynamics of the optical wave~\cite{gouts-pra2018}
\begin{equation}
  \psi(r,z)=
  \frac{\rho A(\rho)e^{i\Psi_0}}{iz}
  \frac{(2\pi k)^{1/2}}
  {\left|\frac1z-\frac1{z_c}\right|^{1/2}}
  J_0\left(\frac{kr\rho}{z}\right).
  \label{eq:j0:sol01}
\end{equation}
In Eq.~(\ref{eq:j0:sol01})
$\Psi_0=\phi(\rho)+k(r^2+\rho^2)/(2z)+\mu\pi/4$,
$\mu = \sgn( 1/z-1/z_c)$, and we have defined
\begin{equation}
  z_c = -k/\phi''(\rho).
  \label{eq:j0:zc}
\end{equation}
In addition, we define the width of the Bessel beam as
\begin{equation}
  W = 2z/(k\rho).
  \label{eq:j0:W}
\end{equation}
After some calculations, we can express Eq.~(\ref{eq:j0:sol01}) in the simpler form
\begin{equation}
  \psi(r,z)=
  (2\pi k\rho\rho'(z))^{1/2}
  \frac{A(\rho)  e^{i\Psi_0}}{i}
  J_0\left(\frac{2r}{W}\right) 
  \label{eq:j0:sol02}
\end{equation}
where we have selected
\begin{equation}
  \rho'(z)>0
  \label{eq:rhop01}
\end{equation}
and thus $\mu=\sgn(\rho'(z))=1$. We relate the full width at half intensity maximum (FWHM) with $W$ via $W_0=w_0 W$, where $w_0$ is the width of the main lobe of $J_0(2 r)$ at half intensity maximum.

The ray picture of the zeroth-order Bessel-like beams consists of rays emitted from expanding concentric circles with different inclinations that intersect along a focal line that passes perpendicularly through the center of the circles.
The condition given by Eq.~(\ref{eq:rhop01}) results to a $1-1$ correspondence between the location of the ray on the input plane and the on-axis focal distance $z$. Furthermore, since $\rho'(z)>0$ then as $\rho$ increases $z$ also increases. The scenario where $\rho'(z)<0$ also provides such a $1-1$ correspondence but, in this case, rays from smaller $\rho$ focus at larger distances $z$. This leads to the problem of ray interference that reduces the quality of the resulting beam. A more convenient form of Eq.~(\ref{eq:rhop01}) is 
\begin{equation}
  W(z) - W'(z)z>0.
  \label{eq:rhop02}
\end{equation}
The on-axis maximum amplitude of the Bessel beam is related to the amplitude on the input plane as
\begin{equation}
  U(z)=(2k\pi\rho\rho'(z))^{1/2}A(\rho).
  \label{eq:j0:U}
\end{equation}

The above calculations can be used to generate Bessel beams with
preassigned
width $W(z)$ and maximum amplitude $U(z)$ as a function of the propagation distance.
Depending on the complexity of $W(z)$ the calculations can be carried out analytically or numerically.

Let us discuss some specific examples. For a power-law beam width
\[
  W(z) = a+bz^c
\]
Eq.~(\ref{eq:rhop02}) is satisfied as long as $a+b(1-c)z^c>0$. In particular, for a linearly varying (increasing or decreasing) beam width ($c=1$) the phase 
$\phi(\rho) = \rho(kb\rho-4)/(2a)$
is single valued provided that $a>0$. 
On the other hand, if $a=0$ Eq.~(\ref{eq:rhop02}) is satisfied when $c<1$. The required phase on the input plane is then given by
$
\phi =   -k\rho^2[(1-c)/(1-2c)]
(2/(kb\rho))^{1/(1-c)}
$
for $c\neq1/2$, whereas for $c=1/2$ we have
$\phi=-4\log(\rho)/(b^2k)$.
In the particular case where $c=-1$ the width is inversely proportional to the propagation distance. These solutions are called pin-like beams and were recently examined in detail in~\cite{zhang-aplp2019}. Note that for $c=-1$ the phase exponent $3/2$ is characteristic of the Airy beam.

\begin{figure}
\centering
\includegraphics[width=\linewidth]{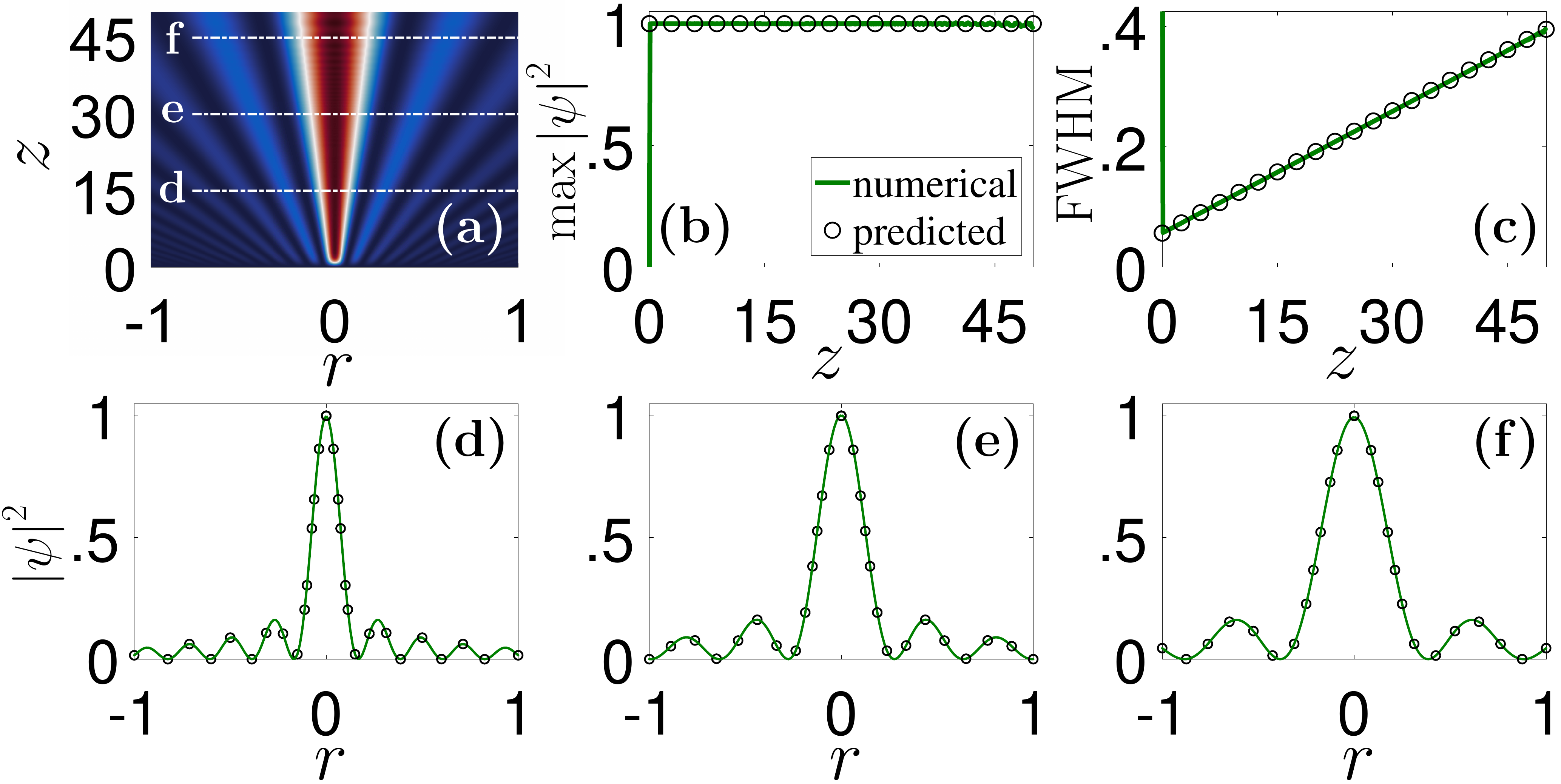} 
\caption{(a) Propagation dynamics of a zeroth order Bessel-like beam. The beam width increases linearly with the propagation distance as $W=a+bz$ with $a=0.05$, $b=0.006$. Also $w_0=1.13$ and $U(z)=1$. In (b) and (c) the maximum intensity and FWHM along $z$ are shown, respectively. (d)-(f) Intensity cross-sections at propagation distances indicated by the white dashed-dotted lines in (a). Numerical (theoretical) results are represented by green solid line (black circles).}
\label{Fig1}
\end{figure}
In all our simulations we normalize the transverse radial coordinate according to $r_0$ and the longitudinal coordinate according to $z_0=kr_0^2$. In addition,
since in many cases $A(\rho)\rightarrow\infty$ as $\rho\rightarrow0$, to eliminate the artificial singularity,
the amplitude on the input plane is multiplied with a hyperbolic tangent function with argument proportional to $\rho$. In Fig.~(\ref{Fig1}) we select the beam width to increase linearly with the propagation distance ($W=a+bz$). We see that there is an excellent agreement between the theoretical and the numerical results. Note that the width of the Bessel beam on the output is $6$ times larger as compared to the initial plane.

In the case of vortex Bessel-like beams, rather than engineering the width of the main (first) lobe of the beam, it is more useful to engineer the inner radius of the cylindrical high intensity surface that surrounds the void region as a function of the propagation distance $R_f(z)$.
In particular, the calculated value of $|\psi(R_f)|$ is selected to be half of the maximum intensity of the first Bessel ring. 
We start by applying a stationary phase approximation to the radial coordinate, $\rho$, of Eq.~(\ref{eq:Fresnel}). As a result we obtain the ray equation
\begin{equation}
  \phi'(\rho)+k(\rho-r\cos s)/z=0
  \label{eq:ray1}
\end{equation}
which is utilized to integrate the Fresnel integral over $\rho$. Subsequently, we apply a stationary phase approximation to the angular variable, $s$, leading to the additional relation for the ray dynamics
\begin{equation}
  n = kr\rho\sin s/z. 
  \label{eq:ray2}
\end{equation}
By directly integrating the Fresnel integral over $s$, we derive the following equation for the amplitude dynamics
\begin{equation}
  \psi(r,\theta,z)=
  \frac{\rho A(\rho) e^{i\Psi_n}}{i^{1+n}z} 
  \frac{(2\pi k)^{1/2}}{\left|\frac1z-\frac1{z_c}\right|^{1/2}}
  J_n\left(\frac{kr\rho}{z}\right), 
  \label{eq:jn:sol01}
\end{equation}
where $\Psi_n = \Psi_0+n\theta$. It is worth pointing out that Eq.~(\ref{eq:jn:sol01}) can be considered as a generalization of Eq.~(\ref{eq:j0:sol01}). The trajectory of a ray on the $r-z$ plane is independent from the launch angle on the input plane. In particular, by eliminating the angle $s$ from the ray Eqs.~(\ref{eq:ray1})-(\ref{eq:ray2}), we find that the rays follow the hyperbolic trajectory
\begin{equation}
  r^2
  =
  \left(\rho+\frac{z}{k}\phi'(\rho)\right)^2
  +\left(\frac{nz}{k\rho}\right)^2.
  \label{eq:ray3}
\end{equation}
We can determine the location of the focal ring from the relation $\dif r^2/\dif z=0$ leading to
\begin{equation}
  (z,r_f) =
  \left(
    -\frac{k\phi'(\rho)\rho^3}{n^2+(\phi'(\rho))^2\rho^2},
    \frac{|n|\rho}{(n^2+(\phi'(\rho))^2\rho^2)^{1/2}}
  \right). 
\end{equation}
Furthermore, by using Eq.~(\ref{eq:ray3}) we can now express $\rho$ as a function of the propagation distance
\begin{equation}
  \rho(z) 
  =
  \left[r_f^2(z)+\left(\frac{nz}{k r_f(z)}\right)^2
  \right]^{1/2}.
  \label{eq:jn:rho}
\end{equation}
The ray picture of the higher-order Bessel-like beams proposed here consists of rays emitted at skewed angles from expanding concentric circles with different inclinations that generate hyperbolic surfaces. The minimum radius of the rays from the axis $r_f$ is achieved at $z_f$.

As in the case of zeroth order Bessel-like beams, we define the width of a vortex Bessel-like beam as
\begin{equation}
  W = \frac{2z}{k\rho}
  =
  -\frac{2\phi'(\rho)\rho^2}{n^2+(\phi'(\rho))^2\rho^2}.
  \label{eq:jn:W}
\end{equation}
The derivative of the phase is related to the vortex trajectory through
\begin{equation}
  \phi'(\rho) = -n^2z/(k\rho r_f^2).
  \label{eq:jn:phip}
\end{equation}
We also satisfy the constraint for an increasing $\rho(z)$ [Eq.~(\ref{eq:rhop02})]. 

Simplified formulas are derived by utilizing the inequality $|\phi'(\rho)|\rho\gg|n|$, which is equivalent to the assumption that the radius of a ray on the input plane is much larger than the focal radius, $\rho\gg r_f$. We would like to point out that this latter approximation is valid in most of the relevant cases. Thus we obtain the following relation
\[
  (z,r_f) =
  (
    -k\rho/\phi'(\rho),-|n|/\phi'(\rho)
  )
\]
for the focal coordinates.
Note that $|\psi(r_f)|$ is in very good agreement with $|\psi(R_f)|$ and thus $r_f\approx R_f$. The largest deviation occurs for $n=1$ ($R_f/r_f\approx0.91$) while for $n=2$ we have $R_f/r_f\approx0.98$. 
In addition the width of the Bessel-like beam takes the simple form
\begin{equation}
  W = -2/\phi'(\rho)=2r_f/|n|.
  \label{eq:jn:W2}
\end{equation}
From Eq.~(\ref{eq:jn:W2}), we see that $W$ and $r_f$ are proportional, as expected due to the self-similar nature of the solutions. Also from Eqs.~(\ref{eq:jn:phip}), (\ref{eq:jn:W2}), we obtain the interesting relationship $r_f=(n/k)(z/\rho)$. It is worth mentioning that, in this approximation, the relation between the width $W$ and the phase $\phi$ is identical to the case of zero vorticity.  
Following the relevant calculations, it can be shown that the beam dynamics is given by 
\begin{equation}
  \psi(r,\theta,z)=
  (2\pi k\rho\rho'(z))^{1/2}
  \frac{A(\rho)e^{i\Psi_n}}{i^{1+n}}
  J_n\left(\frac{2r}{W}\right).
  \label{eq:jn:sol02}
\end{equation}
Equation~(\ref{eq:jn:sol02}), which holds when $|\phi'|\rho\gg|n|$, can be considered as a generalization of Eq.~(\ref{eq:j0:sol02}) for nonzero values of $n$. In order to achieve the intended maximum amplitude $U(z)$ along the propagation distance, we choose the amplitude on the input plane as
\begin{equation}
  A(\rho) = U(z)/[(2\pi k\rho\rho'(z))^{1/2}c_1],
\label{V_Amp}
\end{equation}
where $c_1=\max\left|J_n(r) \right|$.

\begin{figure}
  \centering
  \includegraphics[width=\linewidth]{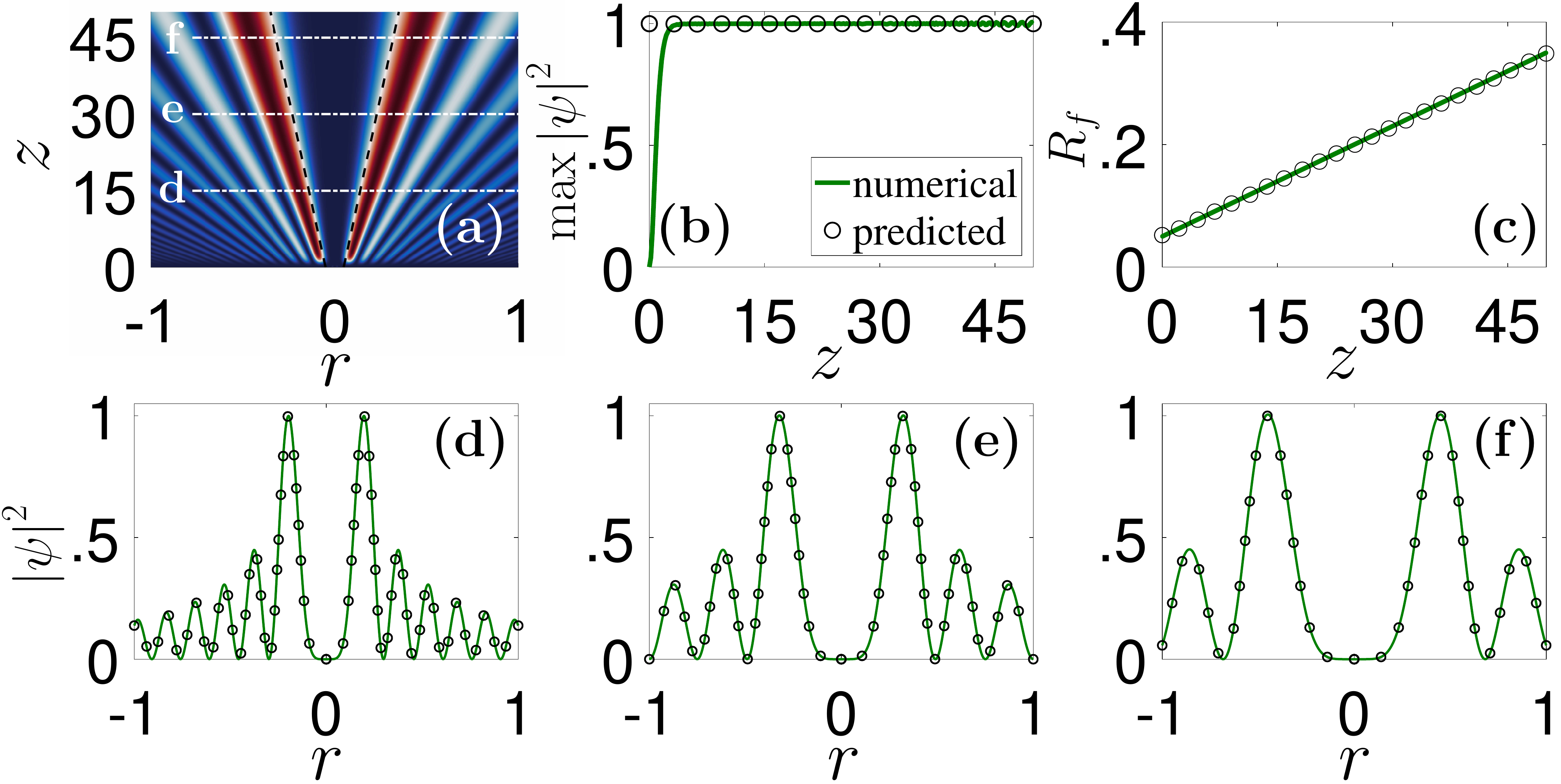} 
  \caption{(a) Propagation dynamics of a self-similar Bessel-like beam carrying orbital angular momentum. The ring radius increases linearly with $z$ as $r_f(z)=a+bz$ with $a=0.05$, $b=0.006$ (black dashed lines). Also $n=3$,
    and $U(z)=1$. In (b) and (c) the maximum intensity and $R_f$ along $z$ are shown, respectively. (d)-(f) Intensity cross-sections at propagation distances indicated by the white dashed-dotted lines in (a). Numerical (theoretical) results are represented by green solid lines (black circles).}
  \label{Fig2}
\end{figure}
We would like to point out that it is possible to extend the propagation distance of these solutions beyond the critical value $z=z_\mathrm{cr}$ after which the solution does not satisfy Eq.~(\ref{eq:rhop01}). In particular, we can select a value $z_m<z_\mathrm{cr}$ such that for $z>z_m$ the Bessel-like beam becomes a regular Bessel beam (having constant width). 

We have performed numerical simulations for different classes of self-similar vortex Bessel-like beams.
In the first example shown in Fig.~\ref{Fig2}, we have selected the vortex radius to increase in a linear manner as $r_f= a+bz$ with positive $a$ and $b$, the maximum amplitude to be constant $U=1$, and the topological charge $n=3$. Interestingly, in accordance to our theoretical model, a $6$ times increment over the initial radius and beam-width is achieved. We see that our theoretical predictions are in excellent agreement with our numerical results. 

\begin{figure}
\centering
\includegraphics[width=\linewidth]{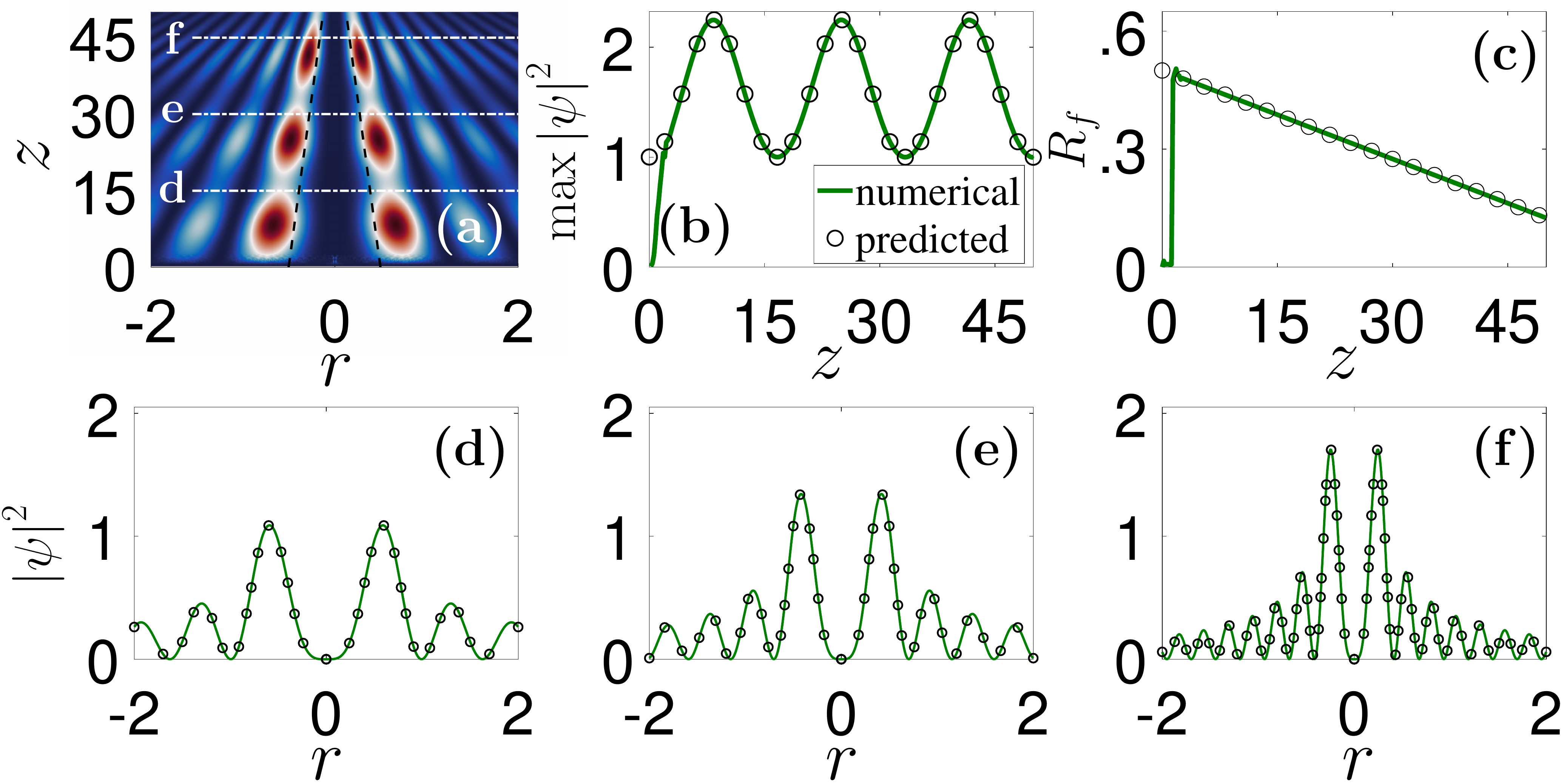} 
\caption{Same as in Fig.~\ref{Fig2} but for linearly decreasing ring radius $r_f=a+bz$  with $a=0.5$, $b=-0.0075$, $n=2$. 
  Here the maximum amplitude is $U(z)=1+0.5 \sin^2(3\pi z/50)$.}
\label{Fig3}
\end{figure}
In the second example shown in Fig.~\ref{Fig3}, we have selected a vortex with topological charge $n=2$, and the core radius to decrease linearly with the propagation distance ($r_f=a+bz$ with $a$ positive and $b$ negative). In addition, the maximum amplitude $U$ is the sum of a constant and a sinusoidal function. During propagation the initial radius decreases by $75\%$. 

\begin{figure}
\centering
\includegraphics[width=\linewidth]{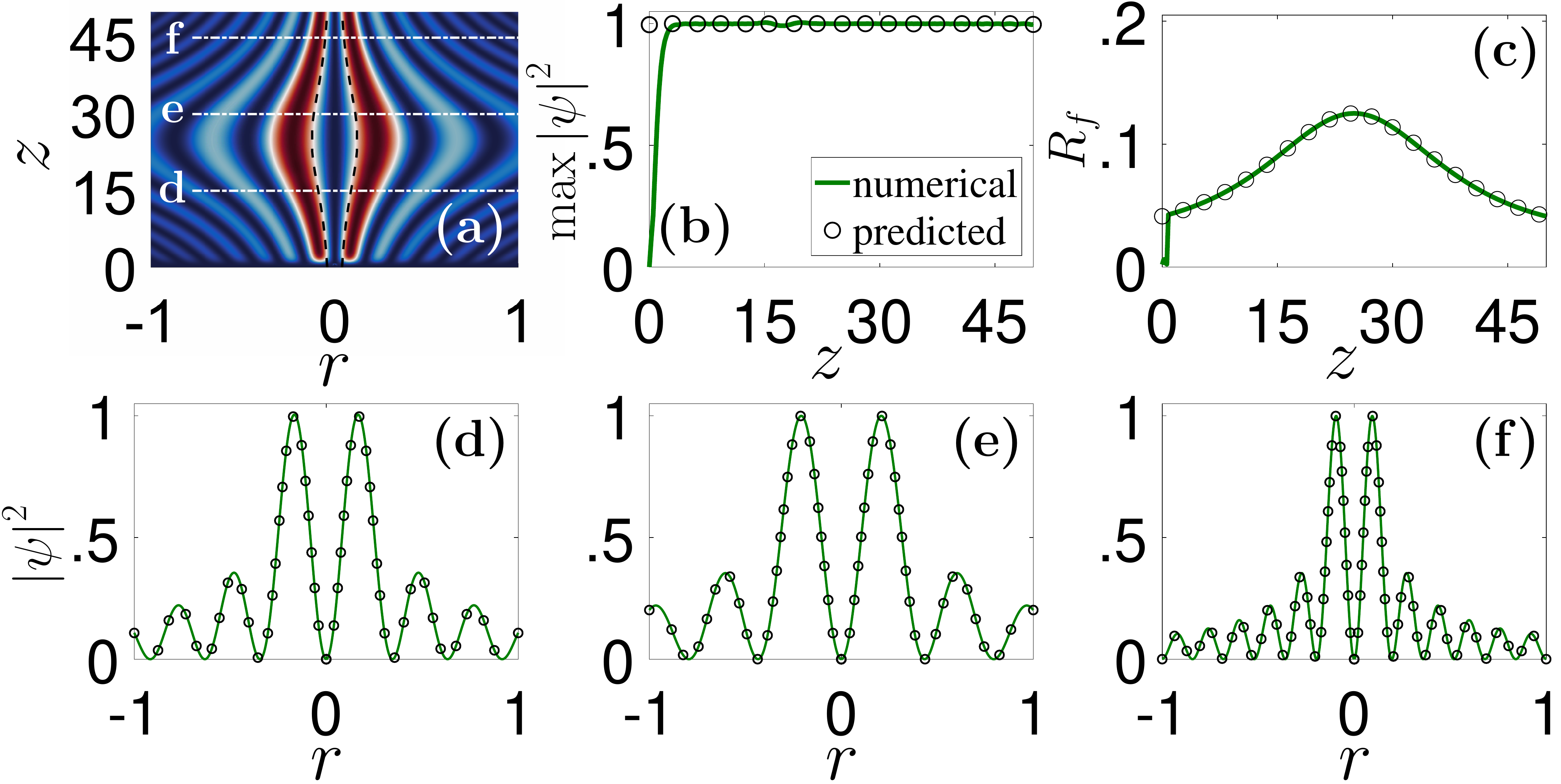}  
\caption{Same as in Fig.~\ref{Fig2}, but for $r_f=a+b\sech[\gamma(z-c)]$ with $a=0.025$, $b=\gamma=0.1$, $c=25$, $n=1$, and $U(z)=1$.}
\label{Fig4}
\end{figure}
In the final example, we have selected a hyperbolic secant modulation of the hollow core radius. In Fig.~\ref{Fig4}, we depict results for the propagation of such a first order vortex Bessel beam, where the maximum intensity is selected to be constant. To ensure that $\rho(z)$ is an increasing function we have tested that Eq.~(\ref{eq:rhop02}) is fulfilled.

In conclusion, we have shown that it is possible to generate self-similar vortex Bessel-like beams with pre-designed dark core radius and fully controllable maximum intensity. In addition, we have used our method to generate Bessel-like beams without vorticity with tunable width and maximum intensity. Our theoretical results are in excellent agreement with our numerical simulations. The ability to dynamically control the parameters of self-healing
Bessel-like beams with or without vorticity is of particular importance, taking into account the applicability of such PI waves in many areas of optics.
Such optical waves can be generated experimentally by encoding both the amplitude and the phase into phase only filters~\cite{lee-ao1979,davis-ao1999}.

\section*{Funding}
This research is funded by the Greek State Scholarships Foundation (IKY), project (MIS-5000432) and by the National Key R\&D
Program of China (Grant No. 2017YFA0303800)

\bigskip

\newcommand{\noopsort[1]}{} \newcommand{\singleletter}[1]{#1}

\renewcommand\refname{FULL REFERENCES}

\end{document}